\documentclass[10pt,aps,pre,twocolumn,groupedaddress,showpacs,showkeys]{revtex4-1}
\usepackage{color}
\usepackage{braket}
\usepackage{graphicx}
\usepackage{amsmath}
\usepackage{amsmath,amssymb,MnSymbol}
\usepackage{IEEEtrantools}

\usepackage{enumitem}
\newcommand{\psitket}{\ensuremath{\ket{\psi,t}}}

\newcommand{\AH}{A_{\mathrm{H}}}
\newcommand{\DHam}{D_{\mathrm{H}}}
\begin{document}

\title{Continuous-Time Quantum Walks on Directed Bipartite Graphs}

\author{Beat T\"odtli}	
\email[]{Corresponding author: beat.toedtli@ffhs.ch}
\author{Monika Laner}
\author{Jouri Semenov}
\author{Beatrice Paoli}
\affiliation{Laboratory for Web Science, Swiss Distance University of Applied Sciences (FFHS)}
\author{Marcel Blattner}
\affiliation{Tamedia Digital Analytics, Tamedia Zurich}
\author{J\'er\^ome Kunegis}
\affiliation{Institute for Web Science and Technologies, University of Koblenz-Landau}
\date{\today}

\begin{abstract}
This paper investigates continuous-time quantum walks on directed bipartite graphs based on a graph's adjacency matrix. 
We prove that on bipartite graphs, probability transport between the two node partitions can be completely suppressed by tuning a model parameter $\alpha$. 
We provide analytic solutions to the quantum walks for the star and circulant graph classes that are valid for an arbitrary value of the number of nodes $N$, time $t$ and the model parameter $\alpha$. We discuss quantitative and qualitative aspects of quantum walks based on directed graphs and their undirected counterparts. 
Numerical simulations of quantum walks on circulant graphs show complex interference phenomena and how complete suppression of transport is achieved near $\alpha=\pi/2$.
By proving two mirror symmetries around $\alpha=0$ and $\pi/2$ we show that these quantum walks have a period of $\pi$ in $\alpha$. We show that undirected edges lose their effect on the quantum walk at $\alpha=\pi/2$ and present non-bipartite graphs that exhibit suppression of transport. Finally, we analytically compute the Hamiltonians of quantum walks on the directed ring graph. 
\end{abstract}


\maketitle

\section{Introduction}
Quantum computing promises to deliver faster computation based on the principles of quantum mechanics.
Shor's \cite{shor94} and Grover's algorithms \cite{grover97} are the most prominent applications of quantum computation, in which a significant speedup over classical computers is demonstrated. The former efficiently factors integers while the latter finds an item in an unsorted database of qubits. 

Quantum information transport across quantum networks is now becoming feasible~\cite{Bao11122012,Duan2001}, and conditions for perfect or pretty good state transfer between nodes on a quantum network are being studied using quantum walks~\cite{Godsil2011,Vinet2012}.
Quantum walks are also used to study the transport of energy excitations. For instance, excitations in light harvesting complexes are transported very efficiently to photosynthetic chemical reaction centers~\cite{Sarovar2010,caruso2009}, promising interesting technological applications such as optimized solar cells.

Quantum walks are used extensively as algorithmic tools for quantum computation. 
They have been defined using two different formulations, discrete-time~\cite{Aharonov2001} and continuous-time~\cite{PhysRevA.58.915,Mulken2011}, with essentially the same computational power~\cite{VenegasAndraca:2012fh} and the latter being a limit case of the former~\cite{Childs2010discCont}. In fact, universal computation is possible using quantum walks on graphs~\cite{Childs2009quantalgo,PhysRevA.81.042330}. 

Farhi and Gutmann \cite{PhysRevA.58.915} investigated the dynamics of continuous-time quantum walks on undirected graphs by introducing quantum systems whose Hamiltonians are based on the adjacency matrix of a graph. Zimbor\'{a}s et al.\ \cite{zimboras13} have started investigations to extend quantum walks on graphs to directed graphs by introducing complex phases. They show that Hamiltonians with a single phase already exhibit a rich phenomenology. In particular, their discovery of complete suppression of transport on the bipartite directed ring graph calls for detailed investigations on this phenomenon in the context of directed quantum walks.

In this paper, we investigate continuous-time quantum walks in XY-spin models based on directed bipartite structures. This structure is present in many systems with a translational symmetry such as one-, two- and three-dimensional lattices, and thus applies to materials as diverse as crystal structures, graphene and carbon nanotubes. Their networked structure can be represented as a directed bipartite circulant graph. In particular, we study quantum walks as a function of the complex phase and compare walks on directed graphs with walks on their undirected counterparts. We focus on the star and circulant graphs, two extreme types of bipartite graphs that differ most notably in their partition size ratios. We simulate quantum walks on circulant graphs using four model graphs, intended to limit the computational complexity of the problem.

\section{Formalism}\label{sec:Methods}
\subsection{Hermitian Hamiltonians of Directed Graphs}
\label{sec:hamiltonians}
In a classical random walk, a walker moves along the edges of a connected graph $\Gamma\left(V,E\right)$ of $N$ nodes, with a hopping probability assigned to each edge $e \in E$ between nodes $i \in V$ and $j \in V$. For directed graphs, edges in $E$ are ordered pairs of nodes. An edge between nodes $i$ and $j$ is called bidirected if both $(i,j)$ and $(j,i)$ are in $E$.  Walks on undirected graphs can be modeled as walks on the corresponding directed graph with bidirectional edges. 

In a quantum walk, a state vector $\psitket$ undergoes a time evolution given by the solution of the Schr\"odinger equation (using $\hbar=1$):
\begin{equation}\label{eq:timeevolU}
 \psitket = e^{-iHt}\ket{\psi,t=0}.
\end{equation}
In the case of an undirected (or bidirected) graph, as studied by Farhi and Gutmann \cite{PhysRevA.58.915}, the Hamiltonian $H$ is set equal to the symmetric adjacency matrix $A$ of the graph.
For directed graphs the adjacency matrix will not in general be Hermitian and thus the time evolution will not be unitary. Therefore this approach does not produce valid Hamiltonians for quantum walks.
To incorporate edge directions into a quantum mechanical Hamiltonian the antisymmetric part of the corresponding adjacency matrix $A$ needs to be included. 
Since a Hermitian Hamiltonian is not necessarily symmetric, the edge directions can be encoded using $N\left(N-1\right)/2$ possible phases~\cite{zimboras13}.
In this paper the investigations are restricted to a single phase denoted by $\alpha$.
Any single-phase Hermitian linear combination of $A$ and $A^{T}$ is proportional to the \emph{Hermitian adjacency matrix}
\begin{equation}\label{eq:AHDef}
\begin{split}
	\AH &= e^{i\alpha}A+e^{-i\alpha}A^T\\
 &=\left(A+A^T\right)\cos\alpha+i\left(A-A^T\right)\sin\alpha.
\end{split}
\end{equation}
The angle $\alpha$ rotates the symmetric and antisymmetric components of the graph. 
Since $\AH\left(\alpha=0\right)=A+A^T= A^\text{sym}$, $\AH$ can be regarded as a generalization of the Hamiltonians studied by Farhi and Gutmann~\cite{PhysRevA.58.915}.

\subsection{Quantum Spin Systems}\label{sec:qss}
Many physical systems can be described as two-level systems in which the relevant physical properties can be described using a two-dimensional Hilbert space. Examples are
two-level atoms, polarized photons and spin-1/2 particles.
Composite systems formed by two level systems are usually termed \emph{spin 
systems}.
In the following we present a spin system that realizes the type of quantum 
walks on directed graphs studied in this article. A Hamiltonian expressed in terms of spin 
 operators might facilitate the experimental realization of the quantum walk 
but trapped ions, nuclear magnetic resonance or other approaches can 
also efficiently run a quantum walk. 
Therefore, our approach applies to any system whose Hamiltonian can be
represented as a matrix in the given form, even if it does not arise from a spin system.   
In fact, any universal quantum computer is able to simulate Hamiltonians with local 
interactions~\cite{10.2307/2899535}.

Models of interacting spin systems can be set up such that they realize Hamiltonians of directed graphs. 
Consider the XY-spin model
\begin{align}
 H &= 2\sum_{\substack{m,n=1\\m\neq n}}^N 
J_{mn}\left[S^{x}_mS^{x}_n+S^{y}_mS^{y}_n\right]
\end{align}
where $S^{x}_m,S^{y}_m,S^{z}_m$ are the spin operator components for particle $m$. $J$ is the exchange interaction between spins. 
By construction, this Hamiltonian commutes with the total spin and the sum of all third spin components, implying 
the conservation of the total spin and total $z$-component of the spin. 
The state space decomposes into a sum of subspaces ($H$-eigenspaces) labeled 
by the quantum numbers of the total spin $s$ and the sum of z-spin 
components $s_z$. 
According to the Schr\"odinger equation the time-evolution of the states is given by
\begin{equation}\label{eq:U}
\ket{s,s_z,t} = U\left(t\right)\ket{s,s_z,t=0} = e^{-iHt}\ket{s,s_z,t=0},
\end{equation}
implying that the quantum numbers $s$ and $s_z$ are conserved under 
time-evolution. 
For example, a single-excitation state evolves into a superposition of states with exactly one spin up and all other spins down.
This state evolution is termed a continuous-time quantum walk. 

In the following, we restrict our attention to single-excitation states 
$\ket{i}$ for $i=1\ldots N$. 
In this subspace the 
Hamiltonian is given by the symmetric matrix
\begin{equation}\label{eq:Hij}
 H_{ij} = \braket{i|H|j}=J_{ij}+J_{ji}.
\end{equation}
Continuous-time quantum walks on graphs are defined by giving a relation 
$J\left(\AH\right)$ between the exchange interaction $J_{ij}$ and the Hermitian 
adjacency matrix of a graph.
For directed graphs, $J\left(\AH\right)$ with $\AH$ as given in 
Eq.~(\ref{eq:AHDef}) provides such a relation that is Hermitian and has a 
non-trivial dependency on the directed nature of the graph.
Our Hamiltonian therefore reads 
\begin{equation}\label{eq:H(J)}
 H = J(\AH) + J^T(\AH).
\end{equation}
The matrix function $J\left(\AH\right) = \sum_{n} j_n \AH^n$ is a general matrix polynomial 
or power series~\footnote{As a consequence of the Cayley-Hamilton theorem, power 
series of matrices (or vector space endomorphisms) form an $N$-dimensional 
vector space and can thus be expressed as a polynomial in $\AH$ of degree at 
most $N-1$.} with coefficients $j_n$.
The case $J\left(\AH\right)=j_1 \AH$ leads to $H=j_1\left(\AH+\AH^T\right)=2j_1\left(A+A^T\right)\cos\alpha$ 
showing that the Hamiltonian only depends on the symmetric part $A+A^T$ in Eq.~(\ref{eq:AHDef}), i.e.\ on the undirected 
graph structure. The $\alpha$-dependence consists of a (graph-independent) factor of $\cos\alpha$ and
can be absorbed into the time scale in the exponent of Eq.~(\ref{eq:timeevolU}).
In order to find non-trivial differences between walks on directed and undirected graphs
nonlinear functions $J\left(\AH\right)$ need to be considered. In this article
analyses are done for a general function $J\left(\AH\right)$, but in Sec.~\ref{sec:CTQRWStar} and~\ref{subsec:ringgraphs} we give results
for test cases such as $J\left(\AH\right)=\exp \AH$.

Our main object of study is the time evolution of the probability distribution
\begin{align}\label{eq:Pit}
P\left(i,t\right)=\left|\braket{i,t|i_0,0}\right|^2=\left|\bra{i}e^{-iHt}\ket{i_0}\right|^2
\end{align}
with $\ket{i,t}$ denoting the quantum state of the $i$-th node at time $t$ and with $P\left(i_0,t=0\right)\neq 0$ only for a single node $i_0$.

\subsection{Bipartite and Circulant Graphs}\label{sec:bipOrCirc}
Bipartite graphs are graphs whose nodes can be partitioned into two disjoint sets (partitions) such that no edge connects two nodes in the same set. 
For example, tree graphs such as the star graph and ring graphs with an even number of nodes are bipartite. 
Choosing a node ordering in which all nodes of one partition are in sequence, the adjacency matrix of a bipartite graph is antidiagonal,
\begin{equation}\label{eq:antiblockdiagonalmatrix}
A^{(N\times N)} = \begin{pmatrix}
     {0}^{(m\times m)} & B_1^{(m\times n)} \\
     B_2^{(n\times m)} & 0^{(n\times n)}
    \end{pmatrix}
\end{equation}
where $B_{1,2}$ are the biadjacency matrices of the directed graph. 
If the graph is undirected, then $B_2$ equals $B_1^T$.

A circulant graph is a graph whose nodes can be ordered in such a way that its adjacency matrix is a circulant matrix~\cite{Gray2006}. 
A circulant matrix $M$ is defined by a vector $\left(m_0,\ldots,m_{N-1}\right)$ and the specification that the $i$-th row is given by $\left(m_0,\ldots,m_{N-1}\right)$ circularly shifted to the right $i$ times, i.e.\ $M_{ij} = m_{(j-i)\mod N}$. 
We denote this $N\times N$-matrix by $M=\left[m_0,\ldots,m_{N-1}\right]_c$.

Any circulant matrix $M$ can be diagonalized using the unitary change of basis $D_M = S^* M S$ with
\begin{equation}\label{eq:Smn}
 S_{mn} = e^{2\pi i m n/N}/\sqrt{N}\quad \text{for } m,n=0,\ldots,N-1.
\end{equation}
We use a similar notation to denote diagonalized circulant Hamiltonians $\DHam$ or adjacency matrices $D_{\mathrm A}$, etc.
For a circulant graph the Hamiltonian matrix $H$ will be circulant since it is the sum of products of circulant matrices. 

Note that the dynamics of continuous-time quantum walks on graphs does not depend on the choice of the vertex ordering. A permutation of the node labels can destroy the circularity of the Hamiltonian matrix. A Hamiltonian matrix of a quantum walk on a circulant graph is circulant only in certain orderings of the vertices. 
An even number of nodes is a necessary condition for a circulant graph to be bipartite.
Otherwise it would contain cycles of odd length. 

\subsection{Bipartite Circulant Graphs}\label{sec:bipAndCirc}
The continuous-time quantum walk Hamiltonian matrix $H$ of a circulant graph is circulant for a general matrix polynomial $J(\AH)$ as long as $A$ is chosen circulant.
Here we focus on circulant graphs that are bipartite. Then their number of nodes $N$ is even (see Sec.~\ref{sec:bipAndCirc}).
Their continuous-time quantum walk Hamiltonian matrix 
$H=\left[h_0,\ldots,h_{N-1}\right]_c$ can 
be expressed as a block structure after a change of basis $\bar{H}=PHP^T$ using 
\begin{align}
P_{ij}&=\begin{cases}
1& \text{if~}\bigl(\lfloor {2i/N}\rfloor-j+2i\bigr)\bmod N=0,\\
0& \text{otherwise,}
\end{cases}
\end{align}
where $i,j=0,\ldots,N-1$ and $\lfloor{.}\rfloor$ denotes the floor function.
The following diagram illustrates the form of $P$ for $N=8$:
\begin{align}
P\left(N=8\right)\sim\,\raisebox{-0.44\height}{\includegraphics[height=2.5cm]{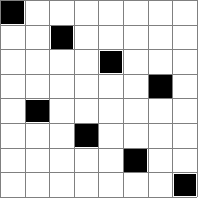}}\:,
\end{align}
with the black squares indicating matrix elements $P_{ij}=1$.
The matrix $\bar{H}$ decomposes into four equal-sized square
blocks,
\begin{align}\label{eq:Hbar}
 \bar{H}=\begin{pmatrix}
  \left[h_{0\phantom{-N}},h_2,\ldots,h_{N-2}\right]_c&  \left[h_1,h_3,\ldots,h_{N-1}\right]_c\\
  \left[h_{N-1},h_1,\ldots,h_{N-3}\right]_c&  \left[h_0,h_2,\ldots,h_{N-2}\right]_c\\
         \end{pmatrix}.
\end{align}
Each block of $\bar H$ is given by a circulant matrix with even- or odd-indexed coefficients only. 
The even-indexed coefficients are present on the block diagonal and the odd-indexed coefficients on the block antidiagonal.

Therefore any circulant matrix $C=\left[c_0,\ldots,c_{N-1}\right]_c$ with $c_{i}=0$ whenever $i$ is even can be transformed into the form of Eq.~(\ref{eq:antiblockdiagonalmatrix}). If in addition $c_{i}\in \{0,1\}$ for all $i$ odd then $C$ viewed as an adjacency matrix defines a bipartite circulant graph. 

\section{Results and Discussion}\label{sec:Results}

\subsection{Suppression of Transport on Bipartite Graphs}\label{sec:suppTransport}
In this section we present our main finding that for $\alpha=\pi/2$, complete suppression of transport occurs for any bipartite graph.

Hamiltonians of the form of Eq.~(\ref{eq:H(J)}) are block diagonal at $\alpha=\pi/2$ if the graph is bipartite. This holds for any power series or polynomial $J\left(\AH\right)=\sum_{n=0}^{N-1} j_n \AH^n$.
For $\alpha=\pi/2$, $\AH\left(\alpha=\pi/2\right)=i\left(A-A^T\right)$ is antisymmetric. It follows that $\AH^T=-\AH$ at $\alpha=\pi/2$ and that odd powers of the Hermitian adjacency matrix $\AH$ cancel in the Hamiltonian: 
\begin{equation}
 \begin{split}
 H\left(\alpha=\pi/2\right) &= J\left(\AH\right)+J^T\left(\AH\right) \\
 &= \sum_{n=0}^{N-1} \left[j_n \AH^n+ j_n \left(\AH^T\right)^n\right]\\
 &=\sum_{n=0}^{N-1}j_n\left(1+\left(-1\right)^n\right)\AH^n\\
 &=2j_0 I+2J_{\text{even}}\left(\AH\right).
 \end{split}
\end{equation}
Here the notations $J_{\text{even}}\left(\AH\right)=\sum_{n\geq 1}j_{2n}\AH^{2n}$ and $I$ for the identity matrix are introduced.
For bipartite graphs and a suitable node labeling, the adjacency matrix $A^{\left(N\times N\right)}$ is block antidiagonal and even powers of $\AH$ are block-diagonal. As a consequence $H\left(\alpha=\pi/2\right)$ is block-diagonal. The state space is therefore split into a sum of two subspaces. Probability transport between the two subsystems (node partitions) is completely suppressed. In physical terms, this corresponds to two isolated subsystems that evolve independently in time. Assuming that the phase $\alpha$ can be tuned experimentally to $\alpha=\pi/2$ (e.g., using magnetic fields), this opens a way to continuously isolate two subsystems that are otherwise strongly coupled.

\subsection{Quantum Walks on Star Graphs}\label{sec:CTQRWStar}
In this section we apply quantum walks to directed and undirected star graphs with $N+1$ nodes.
The adjacency matrices of the directed and undirected star graphs are given by
\begin{align}
A^{\text{dir}}_{ij} &=\delta_{i0}\left(1-\delta_{j0}\right)
= \begin{pmatrix}
                      0&1^{1\times N}\\
                      0^{N\times 1}&0^{N\times N}
	    \end{pmatrix}\quad\text{and} \\
A^{\text{undir}}&=A^{\text{dir}}+A^{\text{dir},T}
= \begin{pmatrix}
                      0&1^{1\times N}\\
                      1^{N\times 1}&0^{N\times N}
	    \end{pmatrix},
\end{align}
with matrix indices $i,j=0,\ldots,N$ and $\delta_{ij}$ being the Kronecker delta.
The index $0$ is assigned to the central node of degree $N$ and indices $1,\ldots,N$ are assigned to the peripheral nodes. 
Focusing on the directed star graph $\AH=A^{\text{dir}}e^{i\alpha}+A^{\text{dir},T}e^{-i\alpha}$, one has 
\begin{align}
\AH &=\begin{pmatrix}
 0&e^{i\alpha}&\cdots&e^{i\alpha}\\
e^{-i\alpha} &0&\cdots&0\\
 \vdots&\vdots&\ddots&\vdots\\
e^{-i\alpha}&0&\cdots&0
\end{pmatrix},
\AH^2
&=\begin{pmatrix}
 N&0&\cdots&0\\
 0&1&\cdots&1\\
 \vdots&\vdots&\ddots&\vdots\\
 0&1&\cdots&1
\end{pmatrix}.
\end{align}
Note that $\AH^2$ is not $\alpha$-dependent. For $n\geq 0$, we find
\begin{align}
\AH^{2n+1}&=N^n \AH\qquad\text{and}\\
\AH^{2(n+1)}&=N^n \AH^2.
\end{align}
Splitting $J\left(\AH\right)$ into constant, even and odd functions of $\AH$, one obtains
\begin{equation}\label{eq:alphadep}
\begin{split}
J\left(\AH\right)=&j_0\AH^0+J_{\text{even}}\left(\AH\right)+J_{\text{odd}}
\left(\AH\right)\\
=&j_0 I+\frac{J_{\text{even}}\left(\sqrt{N}\right)}{N}\AH^2
+\frac{J_{\text{odd}}\left(\sqrt{N}\right)}{\sqrt{N}}\AH
\end{split}
\end{equation}
where $J_{\text{even}}$ and $J_{\text{odd}}$ are even and odd functions of their argument.
Eq.~(\ref{eq:alphadep}) reveals the general structure of the $\alpha$- and $N$-dependence for any regular function $J$. 
The Hamiltonian is
\begin{equation}\label{eq:StarH}
\begin{split}
 H =& J+J^T =2j_0I+ \frac{2}{N}J_{\text{even}}\left(\sqrt{N}\right)\AH^2\\
 &+\frac{2\cos\left(\alpha\right)}{\sqrt{N}}J_{\text{odd}}\left(\sqrt{N}\right) \AH\left(\alpha=0\right).
\end{split}
\end{equation}
This result expresses the Hamiltonian in terms of $\AH$ at $\alpha=0$ and $\AH^2$ only, although the original definition contained an arbitrary power series $J(\AH)$. 
It shows the $\alpha$- and $N$-dependence explicitly, again for a general function $J$. An $\alpha$-dependency is introduced only when $J\left(\AH\right)$ is not an even function. 
At $\alpha=\pi/2$, $H$ is block diagonal since the block antidiagonal term $J_{\text{odd}}\left(\AH\right)$ does not contribute. This demonstrates complete suppression of transport because the Hamiltonian in quantum mechanics is the generator of time translations and a block diagonal Hamiltonian only generates state transitions within each state subspace. 

To study the time dependence at finite $t\neq 0$, we use a specific exchange interaction $J$. We choose $J=\exp(\AH)$ in order to include all powers of $\AH$ (and, by Eq.~(\ref{eq:StarH}), of $\sqrt{N}$) in a non-trivial manner. We find
\begin{align}
j^0&=1\\
 J_{\text{even}}\left(\sqrt{N}\right)&=\cosh\left(\sqrt{N}\right)-1\\
 J_{\text{odd}}\left(\sqrt{N}\right)&=\sinh\left(\sqrt{N}\right).
\end{align}
The time evolution of the initial state $\braket{j|\psi\left(0\right)}=\delta_{j0}$ at $t=0$ is evaluated by computing the matrix exponential $U_{j0}\left(t\right)=\left[\exp\left(-i H t\right)\right]_{j0}$.
The probability for the system to be in state $\ket{i}$ at time $t$ is
\begin{align}
	\left|\braket{i|\psi\left(t\right)}\right|^2&=\left|\braket{i|U\left(t\right)|\psi\left(0\right)}\right|^2=\sum_{j=0}^N\left|U_{ij}\braket{j|\psi\left(0\right)}\right|^2\\
&=\left|U_{i0}\right|^2
= \begin{cases}\label{eq:star_probability}
                                       \frac{1}{2}\left(1+\cos\omega t\right) & i=0,\\
                                       \frac{1}{2N}\left(1-\cos\omega t\right)& i=1\ldots N,
                                       \end{cases}
\end{align}
where the oscillation frequency $\omega$ is
\begin{equation}\label{eq:omegadir}
\omega = 4\sinh\left(\sqrt{N}\right)\cos\alpha.
\end{equation}
These results are valid for an arbitrary number of edges $N$.
Note that $\omega$ is $\alpha$- and $N$-dependent.
A variation in $N$ causes a reweighting of the probabilities of the external nodes shown in Eq.~(\ref{eq:star_probability}) and also affects the oscillation frequency $\omega$. 
The reweighting of the external nodes is a consequence of the star graph's symmetry with respect to permutations of the external nodes. The harmonic oscillation is then essentially the dynamical behavior of a 
2-node line graph, or equivalently a star graph with $N=1$.
The $\sinh\sqrt{N}$-dependency of $\omega$ is a consequence of the choice of $J=\exp\left(\AH\right)=\sinh\left(\AH\right)+\cosh\left(\AH\right)$. 

The results for the undirected star graph are again given by Eq.~(\ref{eq:star_probability}), but with
\begin{equation}\label{eq:omegaundir}
\omega = 8\sinh\left(\sqrt{N}\right)\cos\alpha.
\end{equation}
The oscillation frequencies of the directed and undirected star graph differ by a factor of two.
Thus the time evolution on the undirected star graph is twice as fast as on the directed star graph.

At $\alpha=\pi/2$, the oscillation frequency $\omega$ vanishes both for the directed and undirected star graph (Eqns.~(\ref{eq:omegadir}) and~(\ref{eq:omegaundir}), respectively). 
This implies a complete suppression of transport as predicted in Sec.~\ref{sec:suppTransport} since all directed and undirected star graphs are bipartite.

\subsection{Quantum Walks on Bipartite Circulant Graphs}
In this section we show that for circulant matrices, the diagonalized Hamiltonian $\DHam$ and time evolution operator $D_U$ can be calculated 
analytically for a general matrix polynomial $J\left(\AH\right)$, for a general phase $\alpha$ and a general number of nodes $N$,
because all $N\times N$ circulant matrices are diagonalizable using the change of basis given in Eq.~(\ref{eq:Smn}).

%
If the adjacency matrix $A$ is circulant, then $A^T$, $\AH$, $J$, $H$ and $U$ are circulant as well. 
Thus $\DHam = S^* H S$ is diagonal and gives the spectrum of the quantum system.
Diagonalizing $\AH$ one obtains
\begin{equation}\label{eq:DAH}
  \left[D_{\AH}\left(\alpha\right)\right]_{mm} =2\sum_{k=0}^{N-1}a_k\cos\left(\alpha-\frac{2\pi mk}{N}\right)
\end{equation}
where the $a_k$ denote the entries of the first row of the circulant adjacency matrix $A$.
For circulant matrices a diagonalization of the identity $\AH^T\left(\alpha\right)=\AH\left(-\alpha\right)$ (from Eq.~(\ref{eq:AHDef})) results in 
$D_{\AH^T}\left(\alpha\right)=D_{\AH}\left(-\alpha\right)$. Using this, the diagonal matrix $\DHam$ is obtained as
\begin{align}\label{eq:DH}
\DHam&=S^*HS=  J\left(D_{\AH}\left(\alpha\right)\right)+J\left(D_{\AH}\left(-\alpha\right)\right).
\end{align}
The time evolution operator is calculated using 
\begin{align}\label{eq:USHS}
 U = e^{-iHt}=S \exp\left(-i\DHam t\right)S^*.
\end{align}
This form uses only two matrix multiplications and is therefore rather efficient
for numerical simulations.
Eqns.~(\ref{eq:timeevolU}) and  (\ref{eq:Pit}) then compute the quantum walk 
$P\left(i,t\right)$. 

\subsection{Simulations of Quantum Walks on Bipartite Circulant graphs}\label{subsec:ringgraphs}
The dynamical behavior of a single node in a quantum walk on the undirected ring graph has been studied before~\cite{tsomokos2008state,PhysRevA.73.012313}. Some insight into quantum walks on circulant graphs is found by visually inspecting numerical simulations. In the following we simulate the time- and $\alpha$-dependence of the probability flow $P\left(i,t\right)$ for all graph nodes, varying the parameter $\alpha$ between zero and $\pi / 2$. In Sec.~\ref{sec:CirculantAnalyticResults} several
characteristics of these simulated quantum walks are explained using analytic methods.

\begin{figure*}
  \includegraphics[width=1.0\textwidth]{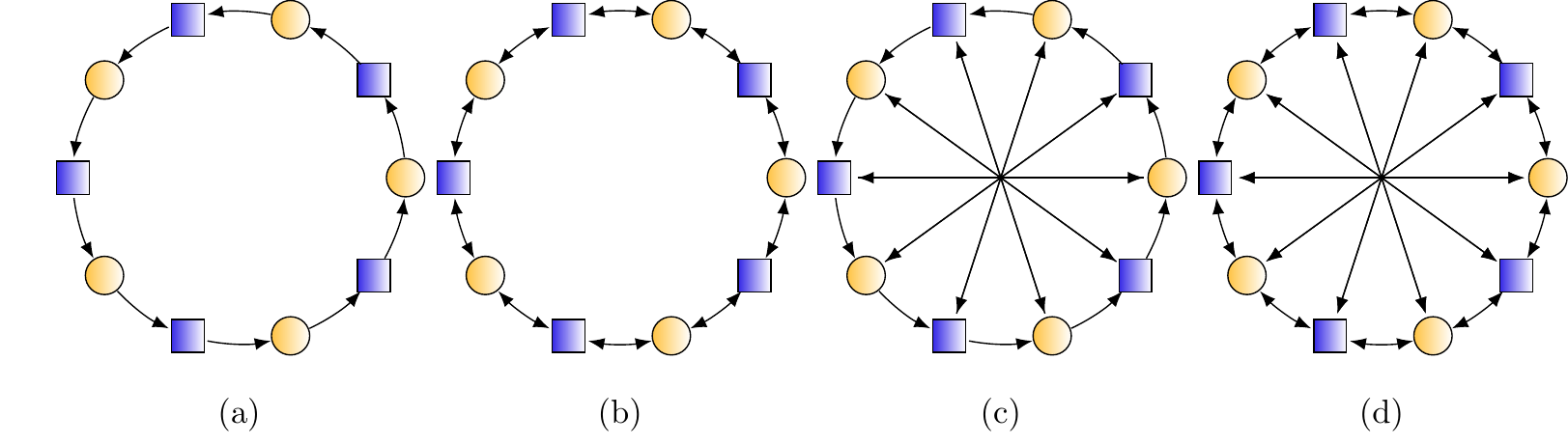}
  \caption{(Color online) Bipartite circulant graphs with $N=10$ nodes. The two node types correspond to the graph's node partitions. The M\"obius ladder graphs (c) and (d) are bipartite if $N/2$ is odd. Continuous-time quantum walks on ring graphs corresponding to (a) and (b) with $N=200$ nodes are shown in Fig.~\ref{fig:ringfigures}. Walks on M\"obius ladder graphs (c) and (d) with $N=202$ nodes are shown in Fig.~\ref{fig:circfigures}.}\label{fig:circfiguresN10}
  \label{fig:circulantGraphs}
\end{figure*}

We study two types of bipartite circulant graphs: ring graphs and directed M\"{o}bius ladder graphs.
M\"{o}bius ladder graphs can be characterized as ring graphs of even length with additional edges connecting every node with the diametrically opposite one~\cite{moebius-ladder} (see Fig.~\ref{fig:circulantGraphs}). They are bipartite only if $N/2$ is odd. 

Continuous-time quantum walks on circulant graphs starting from a uniform initial state are stationary.
Therefore we choose an initial state that is fully localized on one node, $\ket{i,t=0}=\delta_{i,100}$. 
We use natural units, time being measured in units of $1~\text{eV}^{-1}$. 
$J$ is chosen as $J = \exp{\AH}$ in units of $\text{eV}/\hbar$,
and the number of nodes is set to $N=200$ and $N=202$.
Experiments not shown in this paper confirm the general property of circulant Hamiltonians
that the behavior of the system does not qualitatively depend on $N$, except
for scaling effects due to different cycle lengths.

\begin{figure*}
  \includegraphics[width=1.0\textwidth]{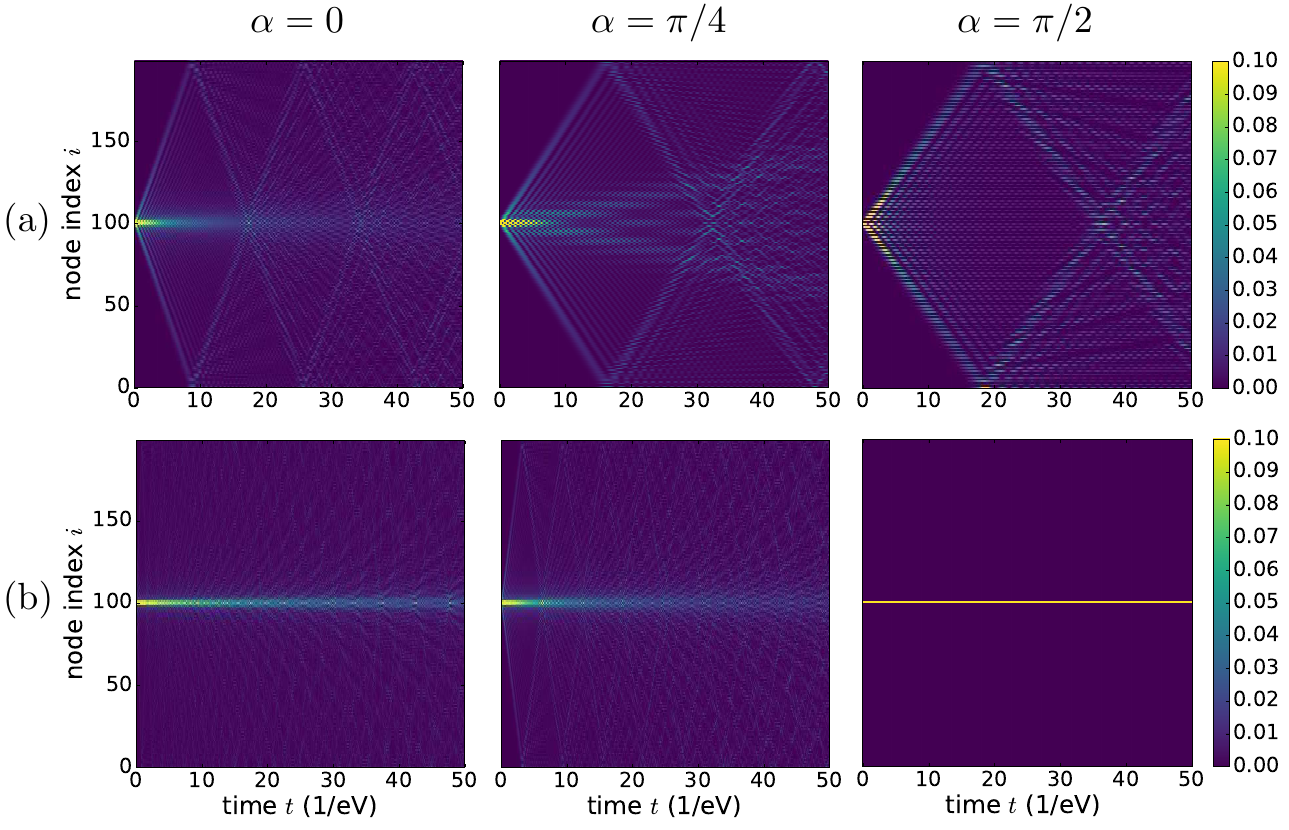}
  \caption{(Color online) Time evolution of the quantum walk on a directed ring (a) and an undirected ring (b) for different values of the phase $\alpha = 0$ (left column), $\alpha = \pi/4$ (middle column), and $\alpha = \pi/2$ (right column). Both rings have $N=200$ nodes. At $t=0$, the quantum walk is started at node $i_0=100$.}
  \label{fig:ringfigures}
\end{figure*}

\subsubsection{Directed and Undirected Ring Graphs}
Fig.~\ref{fig:ringfigures} shows the continuous-time quantum walk on the directed (upper row) and undirected (lower row) ring graph. The corresponding graphs are illustrated in Fig.~\ref{fig:circulantGraphs}a and Fig.~\ref{fig:circulantGraphs}b, respectively, but with $N=10$ nodes for sake of clarity. 

A series of oscillating waves leaves the initial probability concentration at $i_0=100$ in a first wavefront. 
Both for the directed and undirected ring (Figs.~\ref{fig:ringfigures}a and \ref{fig:ringfigures}b)
their velocities decrease from $\alpha=0$ to $\alpha=\pi/4$, where they travel around half of the ring in $9~\text{eV}^{-1}$ to $17~\text{eV}^{-1}$. For the directed ring, the velocity continues to decrease from $\alpha=\pi/4$ to $\alpha=\pi/2$ at a lower rate. In the case of the undirected ring in Fig.~\ref{fig:ringfigures}b, propagation continuously slows down and comes to a complete stop at $\alpha=\pi/2$. 

At $\alpha=0$, for both the directed and undirected ring most of the probability stays around node $i_0=100$. 
For the directed ring, the central probability concentration at node $i_0=100$ is less pronounced and disperses in a secondary wavefront for $\alpha\gtrsim\pi/4$. 
Its propagation velocity reaches its maximum at $\alpha=\pi/2$ where it equals the propagation velocity of the first wavefront.  
The suppression of transport is visible (in Fig.~\ref{fig:ringfigures}a, right panel) as horizontal dark lines interrupting the wavefront and corresponding to a probability value of zero on all the odd-valued node indices, $P\left(i~\mathrm{odd},t\right)=0$. The features described here are discussed further in Sec.~\ref{sec:CirculantAnalyticResults}.

\subsubsection{M\"obius Ladder Graphs}
\begin{figure*}
  \includegraphics[width=1.0\textwidth]{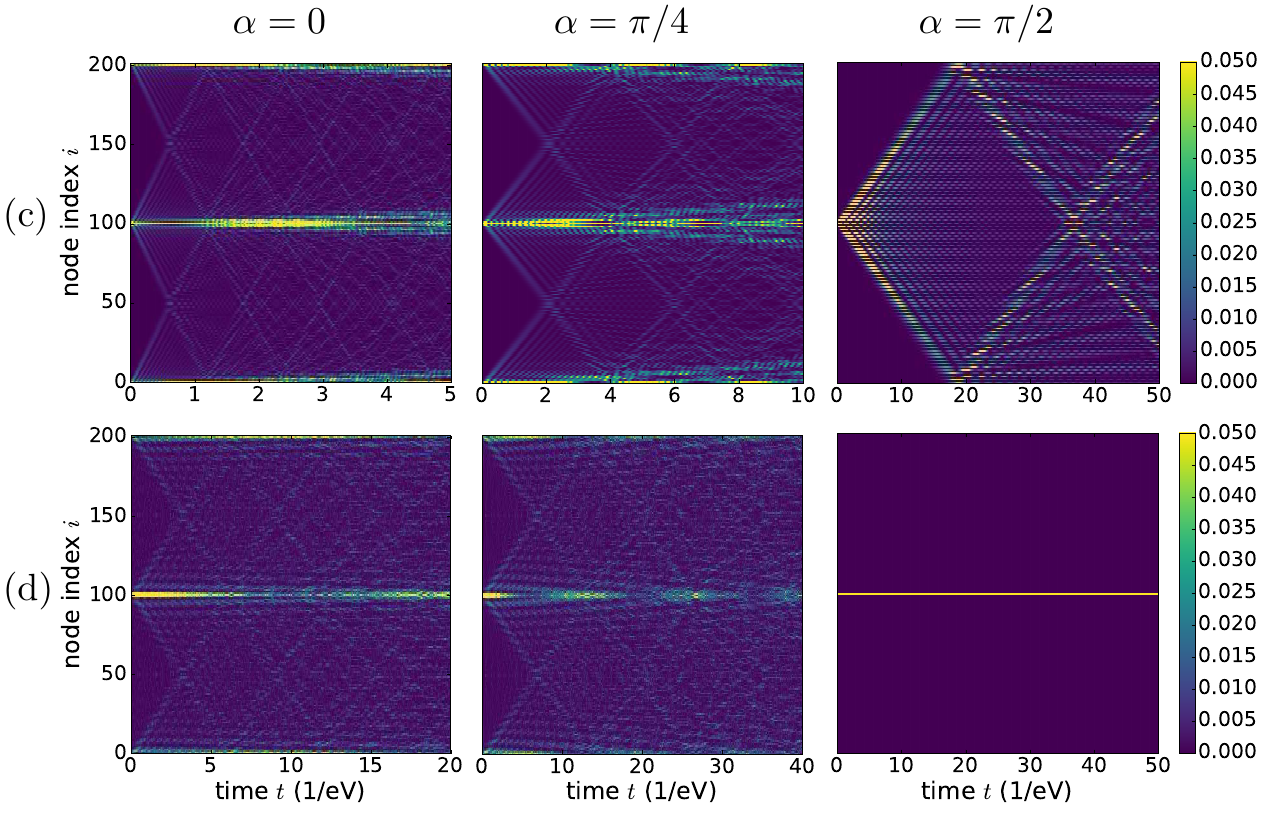}
  \caption{(Color online) Time evolution of quantum walks on directed and undirected M\"obius ladder graphs as shown in Figs.~\ref{fig:circfiguresN10}c and \ref{fig:circfiguresN10}d, but with $N=202$ nodes. The walks start at node $i_0=100$. The values of the phase $\alpha$ are $\alpha = 0$ (left column), $\alpha = \pi/4$ (middle column), and $\alpha = \pi/2$ (right column). 
  }
  \label{fig:circfigures} 
\end{figure*}

Fig.~\ref{fig:circfigures} shows the continuous-time quantum walk on M\"obius ladder graphs as shown in Fig.~\ref{fig:circulantGraphs}c and Fig.~\ref{fig:circulantGraphs}d, but with $N=202$ nodes. They differ from the ring graphs by the presence of additional links $i\rightarrow i+N/2$.
These walks are similar to the ones on ring graphs (see Fig.~\ref{fig:ringfigures}), but with notable differences. 
The probability concentration around $i_0=100$ and the first wavefront can be observed for both graphs. For $\alpha\neq \pi/2$, the links $i\rightarrow i+N/2$ lead to probability wavefronts and a central beam based around the opposite node at index $i=201$. 
The propagation of the first wavefront around the ring proceeds much faster, with the wavefront reaching the opposite node at $\alpha=0$, $\alpha=\pi/4$, and $\alpha=\pi/2$ in $1~\text{eV}^{-1}$, $4~\text{eV}^{-1}$, and $19~\text{eV}^{-1}$ for the directed outer ring, and at $\alpha=0$, $\alpha=\pi/4$ in $6~\text{eV}^{-1}$ and $12~\text{eV}^{-1}$ for the undirected outer ring, respectively.  
The wavefronts slow down as $\alpha$ is varied from $0$ to $\pi/2$, but the effect is much more pronounced for the directed outer ring (Fig.~\ref{fig:circfigures}c) than for the directed ring graph of Fig.~\ref{fig:ringfigures}a. 
Near $\alpha=\pi/2$, the slow propagation of the directed ring is reached.
The split of the central beam into a secondary wavefront can be seen for the directed outer ring (Fig.~\ref{fig:circfigures}c), but at $\alpha=\pi/4$ its velocity is much slower than for the directed ring (Fig.~\ref{fig:ringfigures}a). 
As in the case of the directed ring graph the interference of the first and the second wavefront leaving the initial node happens at $\alpha=\pi/2$ where a complete suppression of transport occurs.
For the undirected outer ring (Fig.~\ref{fig:circfigures}d) the continuous-time quantum walk does not leave the initial node.

\subsection{Analytic Results on Quantum Walks on Bipartite Circulant Graphs}\label{sec:CirculantAnalyticResults}
The continuous-time quantum walks on circulant graphs discussed in the previous section show a rich set of features. 
We explain some of them based on analytic arguments and state the class of directed quantum walks onto which our reasoning extends.
\subsubsection{Suppression of Transport Due to Bipartitivity}
\label{Sec:suppressionOfTransportDueToBipartitivity}
At $\alpha=\pi/2$, quantum walks on all the circulant graphs (a)--(d) discussed in Sec.~\ref{subsec:ringgraphs} show a suppression of transport. If the initial state is localized on the node partition labeled by even indices this suppression of transport takes the form 
$P\left(i,t\right)=0\:\forall t$ if $i$ is odd. It is due to our main result that all bipartite graphs show a suppression of transport (see Sec.~\ref{sec:suppTransport}) and the fact that the graphs (a)--(d) are bipartite. This fact is established in Sec.~\ref{sec:bipAndCirc} where it was shown that if the adjacency matrix of a circulant graph has the form $A=\left[0,a_1,0,\ldots,a_{N-1}\right]_c$, then the graph is bipartite. All graphs in Fig.~\ref{fig:circfiguresN10} are of this form, and their Hamiltonians $H$ will therefore exhibit 
complete suppression of transport at $\alpha=\pi/2$ for any polynomial $J\left(\AH\right)$.
 
\subsubsection{Trivial Time Evolution for Quantum Walks on Symmetric Graphs and Edges}
The graphs (b) and (d) in Fig.~\ref{fig:circfiguresN10} are symmetric graphs. 
Their quantum walks stop evolving at $\alpha=\pi/2$, i.e.\ $P\left(i,t\right)=P\left(i,t_0\right) \forall t$ (see Figs.~\ref{fig:ringfigures}b and \ref{fig:circfigures}d).
This is due to $\AH\left(\alpha=\pi/2\right)=i\left(A-A^T\right) = 0$ because the adjacency matrix $A=A^T$ is symmetric.
The Hamiltonian therefore is diagonal, $H\left(\alpha=\pi/2\right)=2J\left(0\right)=2j_0$. The state evolution is $\ket{\psi,t}=\exp\left( -2ij_0 \left(t-t_0\right)\right)\ket{\psi,t_0}$ and the quantum walk $P\left(i,t\right)=\left|\braket{i|e^{-2ij_0\left(t-t_0\right)}|i}\right|^2=P\left(i,t_0\right)$ evolves trivially. 

The expression $\AH\left(\alpha=\pi/2\right)=i\left(A-A^T\right)$ holds generally and offers insight into why
the quantum walk of Fig.~\ref{fig:ringfigures}a is identical to the one in Fig.~\ref{fig:circfigures}c.
Quantum walks at $\alpha=\pi/2$ depend via $H\left(\AH\right)$ on $A-A^T$ only and not on $A$ and $A^T$ separately.
A symmetric edge between 
nodes $i$ and $j$ has $A_{ij}=A_{ji}=1$ and therefore $\AH\left(\alpha=\pi/2\right)$ and $H\left(\AH\right)$ do not depend on it. Thus the two quantum walks are exactly equal.

Suppression of transport also occurs for non-bipartite graphs if all edges linking nodes within a partition are undirected. These edges cancel in $\AH\left(\alpha=\pi/2\right)$, leaving a continuous-time quantum walk on a bipartite graph that exhibits complete suppression of transport. An example is a directed ring graph with $N=4n$ for any integer $n$ and with additional undirected edges linking opposite nodes.

\subsubsection{Periodicity in the Complex Phase}
Figs.~\ref{fig:ringfigures} and \ref{fig:circfigures} show quantum walks for $\alpha=0$, $\pi/4$ and $\pi/2$ only because apart from the trivial $2\pi$-periodicity, their $\alpha$-dependence has mirror symmetries around $\alpha=0$ and $\pi/2$. 
We prove the mirror symmetry around $\alpha=0$ to be exact for all directed quantum walks with Hamiltonians of the form $H=J\left(\AH\right)+J^T\left(\AH\right)$, i.e.\ for any node $i$ and time $t$ and for all coupling functions $J\left(\AH\right)$. The mirror symmetry around $\alpha=\pi/2$ holds for a large class of quantum walks on bipartite circulant graphs as specified below.

Denote by $P_\alpha\left(i,t\right)$ the explicit $\alpha$-dependence of the quantum walk $P\left(i,t\right)$. Let $\Delta \alpha$ denote the deviation of $\alpha$ from $\pi/2$ by defining $\alpha^\pm=\pi/2\pm\Delta \alpha$. In this notation the symmetries to be shown read 
\begin{align}
 P_\alpha\left(i,t\right)&=P_{-\alpha}\left(i,t\right)\:\forall\alpha,i,t\label{eq:Sym1}\\
 P_{\alpha^+}\left(i,t\right)&=P_{\alpha^-}\left(i,t\right)\:\forall\Delta\alpha,i,t\label{eq:Sym2}
\end{align}
It follows that any quantum walk $P_\alpha\left(i,t\right)$ with these symmetries has a period of (at most) $\pi$ in $\alpha$ and that by defining it on an interval of length $\pi/2$ the symmetries
extend the definition to all real values of $\alpha$.

Eq.~(\ref{eq:Sym1}) is shown by noting that by Eq.~(\ref{eq:AHDef}) it holds that $\AH^T\left(\alpha\right)=\AH\left(-\alpha\right)$ and consequently we have
$H\left(\alpha\right)=J\left(\alpha\right)+J\left(-\alpha\right)$. From this the mirror symmetries $H\left(\alpha\right)=H\left(-\alpha\right)$ and Eq.~(\ref{eq:Sym1}) follow immediately. 

To prove Eq.~(\ref{eq:Sym2}), let $A=\left[a_0,\ldots,a_{N-1}\right]_c$ be the circulant adjacency matrix of a graph with $a_i=0\:\forall i$ even, and let $N$ be even. Then the graph is bipartite due to the results of Sec.~\ref{sec:bipAndCirc}. 
In Appendix~\ref{sec:appendixA} we derive a mirror symmetry in $\alpha$ between energy levels. It reads
\begin{equation}\label{eq:nodeShift}
 \left[\DHam\left(\alpha^+\right)\right]_{mm} = \left[\DHam\left(\alpha^-\right)\right]_{m\pm\frac{N}{2},m\pm\frac{N}{2}}
\end{equation}
with the plus-minus signs suitably chosen such that the matrix indices lie in the range $0,\ldots,N-1$.
Using Eqns.~(\ref{eq:Smn},\ref{eq:DH},\ref{eq:USHS}), Eq.~(\ref{eq:nodeShift}) is written in the form
\begin{equation}
 H(\alpha^+)_{ij}=(-1)^{i+j}H(\alpha^-)_{ij}\label{eq:HUshifted1}.
\end{equation}
For $U=\exp\left(-iHt\right)$ it follows that
\begin{equation}
 U(\alpha^+)_{ij}=(-1)^{i+j}U(\alpha^-)_{ij}\label{eq:HUshifted2}.
\end{equation}
Details are given as in Appendix~\ref{sec:appendixA}.
If we now consider an initial state $\ket{\psi^\text{even}}$ that is localized on even sites only (such that $\braket{i|\psi^\text{even},t=0}=0$ for all odd $i$) we find $\braket{i|\psi^\text{even},t}_{\alpha^-}=\left(-1\right)^i\braket{i|\psi^\text{even},t}_{\alpha^+}$.
Therefore the two quantum walks with phases $\alpha^+$ and $\alpha^-$ are identical:
\begin{align}
\left| \braket{i|\psi^\text{even},t}_{\alpha^-}\right|^2 &= \left| \braket{i|\psi_0^\text{even},t}_{\alpha^+}\right|^2\quad\text{or}\label{eq:probabilitySymmetry1}\\
P_{\frac{\pi}{2}-\Delta\alpha}\left(i,t\right)&=P_{\frac{\pi}{2}+\Delta\alpha}\left(i,t\right)\:\forall \Delta\alpha,i,t.\label{eq:probabilitySymmetry2}
\end{align}
Eqns.~(\ref{eq:probabilitySymmetry1}, \ref{eq:probabilitySymmetry2}) show the mirror symmetry of quantum walks on bipartite circulant graphs in $\alpha$ around $\alpha=\pi/2$.  
\subsubsection{Graph-Locality of the Quantum Walk and Constancy of the Speed of Propagation}
In the quantum walks discussed in Figs.~\ref{fig:ringfigures} and \ref{fig:circfigures}, shortly after $t=0$ the probability distribution $P\left(i,t\lesssim 1\right)$ is mostly localized around node $i_0=100$. Nodes at a certain geodesic graph distance from $i_0$ are only weakly excited before a wave of significant probability amplitude reaches it. The wave travels at a constant group velocity.

The constancy of these propagation velocities is an immediate consequence of the
circularity of the graph: by cyclic permutation invariance all nodes react equally to being excited and pass on excitations at the same rate. The cyclic permutation symmetry $i\rightarrow i+\Delta i$ of the graph is broken by the localized initial state, but the mirror symmetry $H^T=H$ is retained which dictates $P\left(i_0+i,t\right)=P\left(i_0-i,t\right)\:\forall i=0,\ldots,N-1,\forall t$ with all index computations to be taken modulo $N$.

The locality in excitation propagation is not expected to be a general feature but depends on the coupling function $J\left(\AH\right)$. It holds for $J\left(\AH\right)=\sum_p j_p \AH^p$ which have coefficients $j_p$ that decay quickly with $p$. We show that in this case the Hamiltonian is nonzero mainly along or near the diagonal. Thus localized wave functions $\braket{i|\psi,t}=\delta_{i,i_0}$ mainly propagate into their local neighbourhood. 

We show this for the directed ring graph by computing
$H=S\DHam S^*$ for a general matrix polynomial $J\left(\AH\right)$:
\begin{equation}\label{eq:HmnCirc}
 H_{mn}=2\cos\left[\left(m-n\right)\alpha\right]\sum_{p=1}^Nj_p\sum_{k=0}^p\binom{p}{k}\, \delta_{2k,m-n+p}.
\end{equation}
Details are given in Appendix~\ref{sec:appendixB}.
For $j_p=\frac{1}{p!}$, i.e. $J\left(\AH\right)=\exp\left(\AH\right)$, Eq.~(\ref{eq:HmnCirc}) contains 
the binomial factor $j_p\binom{p}{k}=\frac{1}{k!\left(p-k\right)!}$ that peaks at $k=p/2$ (for $p$ even).
The estimate $k\sim p/2$ implies $m\sim n$ due to the Kronecker delta factor $\delta_{2k,m-n+p}$ in Eq.~(\ref{eq:HmnCirc}). The elements $H_{mn}$
are therefore largest near the matrix diagonal. The Hamiltonian $H$ is broadly diagonal, with the magnitude of its matrix elements declining with a growing distance $\left|m-n\right|$ from the diagonal. 

Focusing on $j_p=\delta_{p1}$, i.e.\ $J\left(\AH\right)=\AH$ the Hamiltonian simplifies to 
\begin{align}
 H_{mn}&=2\left(\delta_{m,n-1}+\delta_{m,n+1}\right)\cos\alpha\\
 H&=2\left(A+A^T\right)\cos\alpha\label{eq:HamPPT}
\end{align}
with $A=\left[0,1,0,\ldots,0\right]_c$, indicating that the quantum walk is strictly local. For a localized initial state,
only the neighbouring nodes are affected at any given time. 
Higher powers of $\AH$ (larger $p$) contribute in a less local way. In a qualitative way this can be understood from the fact~\cite{Newman_2010} that classically, the $p$-th power
of the adjacency matrix $\left[A^p\right]_{ij}$ describes the total number of paths of length $p$ from node $j$ to node $i$.

\section{Conclusions and Outlook}
We studied continuous-time quantum walks on directed bipartite graphs by setting up an XY-spin model whose exchange interaction depends on the adjacency matrix of the graph, while hermiticity of the Hamiltonian was ensured by introducing a single complex phase $\alpha$. We prove that complete suppression of transport occurs at $\alpha=\pi/2$ for all such quantum walks on any bipartite graph. 
Our results show that the complex phase $\alpha$ provides a switch to 
isolate the two node partitions from each other and to switch undirected edges 
on and off. This switch is available for bipartite graphs with an arbitrary 
topology and number of nodes $N$.

We provide analytical results for quantum walks on two classes of graphs, star graphs and circulant graphs.
Star graphs are sufficiently simple for the analytic solution of their quantum walks to provide a rather complete picture.
Our analytical results on circulant graphs give access to their accurate numerical simulation. 

Simulations of walks on circulant graphs reveal a rich dynamical structure. 
The suppression of bidirected edges and the mirror symmetries and periodicity in their $\alpha$-dependence are discussed in a rigorous manner. The degree to which some walks are local is analyzed on the ring. 
Further phenomena remain to be investigated such as the secondary wavefront visible at $\alpha\gtrsim \pi/4$, its velocity and interference with the fastest-travelling wave precisely at $\alpha=\pi/2$ and the degree to which the probability accumulation remains concentrated around the initial node. 

The influence of the graph structure and its directionality and of the complex phase $\alpha$ and coupling function $J\left(\AH\right)$ on quantum walks as well as interference effects due to nonlocal initial states need to be investigated further.
Progress in this direction is needed in order to facilitate the engineering of directed quantum walks according to predefined goals and experimental constraints. 

\appendix
\section{Derivation of Eqns.~(\ref{eq:nodeShift}--\ref{eq:HUshifted2})}\label{sec:appendixA}
Let $N$ be an even positive integer and $A$ a 
circulant $N\times N$-matrix of the form
\begin{equation}\label{eq:AHcircbipartite}
A=\left[0,a_1,0,a_3,\ldots,0,a_{N-1}\right]_c.
\end{equation}
Then $\AH\left(\alpha\right)$ and $H\left(\AH\right)$ in Eq.~(\ref{eq:H(J)})
are also circulant. 
Let $\DHam=S^*HS$ denote the diagonal matrix obtained using the unitary change of basis 
given by Eq.~(\ref{eq:Smn}), and 
$\alpha^\pm=\pi/2\pm\Delta\alpha$.
Eq.~(\ref{eq:nodeShift}) states that for any value of $\Delta\alpha$,
\begin{equation}\label{eq:identity}
 \left[\DHam\left(\alpha^+\right)\right]_{m,m} = 
\left[\DHam\left(\alpha^-\right)\right]_{m+\frac{N}{2},m+\frac{N}{2}}
\end{equation}
with the indices $m+N/2$ understood to be taken modulo $N$, as in all subsequent expressions.
It is shown by diagonalizing Eq.~(\ref{eq:H(J)}) and establishing the identity
\begin{equation}\label{eq:relationDAH}
\left[D_{\AH}\left(\pm\alpha^+\right)\right]_{m,m}=\left[D_{\AH}
\left(\mp\alpha^-\right)\right]_{m+\frac{N}{2},m+\frac{N}{2}}.
\end{equation}
From Eq.~(\ref{eq:DAH}) we see that 
\begin{align}
\left[D_{\AH}\left(\pm\alpha^+\right)\right]_{m,m}&=2\sum_{k=0}^{N-1}
a_k\cos\left(\pm\alpha^+ -\frac{2\pi m k}{N}\right)\\
=2\sum_{\substack{k=0\\k~\text{odd}}}^{N-1}a_k&\cos\left(\pm\alpha^+ -\frac{2\pi 
\left(m+\frac{N}{2}\right) k}{N}+\pi k\right)
\end{align}
The summation is restricted because by Eq.~(\ref{eq:AHcircbipartite}), $a_k=0$ whenever $k$ is even. Since $k$ is 
odd, we substitute $\pi k\rightarrow\mp\pi$ in the third term since the 
difference is a multiple of $2\pi$. Noticing that 
$\pm\alpha^+\mp\pi=\mp\alpha^-$ one obtains
\begin{align}
\left[D_{\AH}\left(\pm\alpha^+\right)\right]_{m,m}&=
2\sum_{k=0}^{N-1}a_k\cos\left(\mp\alpha^- -\frac{2\pi \left(m+\frac{N}{2}\right) 
k}{N}\right)\nonumber\\
&=\left[D_{\AH}\left(\mp\alpha^-\right)\right]_{m+\frac{N}{2},m+\frac{N}{2}}
\end{align}
The last equality is due to Eq.~(\ref{eq:DAH}). This 
establishes Eq.~(\ref{eq:relationDAH}).

Eq.~(\ref{eq:identity}) is then obtained by diagonalizing Eq.~(\ref{eq:H(J)}) and using $\AH^T\left(\alpha\right)=\AH\left(-\alpha\right)$,
\begin{align}
\DHam\left(\alpha^\pm\right) &= J\left(D_{\AH}\left(\alpha^\pm\right)\right)+ 
J\left(D_{\AH^T}\left(\alpha^\pm\right)\right)\\
&= J\left(D_{\AH}\left(\alpha^\pm\right)\right)+ 
J\left(D_{\AH}\left(-\alpha^\pm\right)\right).
\end{align}
Switching to component notation we 
have 
\begin{multline}\label{eq:DHJDAH2}
\left[\DHam \left(\alpha^-\right)\right]_{m+\frac{N}{2},m+\frac{N}{2}} 
=J\left(\left[ D_{\AH}\left(\alpha^-\right)\right]_{m+\frac{N}{2},m+\frac{N}{2}}\right) \\
+ J\left(\left[D_{\AH}\left(-\alpha^-\right)\right]_{m+\frac{N}{2},m+\frac{N}{2}}\right).
\end{multline}
and using Eq.~(\ref{eq:relationDAH})
\begin{multline}\label{eq:DHJDAH1}
\left[\DHam \left(\alpha^+\right)\right]_{m,m} = 
J\left(\left[D_{\AH}\left(\alpha^+\right)\right]_{m,m}\right) 
\\
+ 
J\left(\left[D_{\AH}\left(-\alpha^+\right)\right]_{m,m}\right) \\
=
J\left(\left[ D_{\AH}\left(-\alpha^-\right)\right]_{m+\frac{N}{2},m+\frac{N}{2}} \right)
\\
+ 
J\left(\left[D_{\AH}\left(\alpha^-\right)\right]_{m+\frac{N}{2},m+\frac{
N}{2}} \right)
\end{multline}
The two right-hand sides of Eq.~(\ref{eq:DHJDAH2}) and (\ref{eq:DHJDAH1}) are 
equal, which proves Eq.~(\ref{eq:identity}).

By transforming Eq.~(\ref{eq:identity}) back into a relation between the 
circulant Hamiltonians one proves Eqns.~(\ref{eq:HUshifted1},\ref{eq:HUshifted2}). 
Applying $H=S\DHam S^*$ and using $S_{mn}=\left(-1\right)^m S_{m,n+N/2}$ one obtains
\begin{align}
H_{mn}\left(\alpha^+\right)&=\sum_{k=0}^{N-1}S_{mk}\DHam\left[
\left(\alpha^+\right)\right]_{kk}S^*_{kn}\\
=\sum_{k=0}^{N-1}&\left(-1\right)^{m+n} 
S_{m,k+\frac{N}{2}}\DHam\left[\left(\alpha^-\right)\right]_{k+\frac{N}{2},k+\frac{
N}{2}}S^*_{k+\frac{N}{2},n}\nonumber\\
=\sum_{k=0}^{N-1}&\left(-1\right)^{m+n} 
S_{mk}\DHam\left[\left(\alpha^-\right)\right]_{kk}S^*_{kn}\\
=(-1&)^{m+n} H_{mn}\left(\alpha^-\right).
\end{align}
This relation also holds for products of Hamiltonians, e.g.
\begin{equation}
\left[H^2\left(\alpha^+\right)\right]_{mn}=\left(-1\right)^{m+n}\left[
H^2\left(\alpha^-\right)\right]_{mn}
\end{equation}
such that it also holds for the time evolution operator 
$U=\exp\left(-iHt\right)$.

\section{Computation of the Ring Hamiltonian}\label{sec:appendixB}
We compute the Hamiltonian matrix $H=S\DHam S^*$ explicitly for the directed ring graph 
with adjacency matrix $A=\left[0,1,0,\ldots,0\right]_c$. 
The diagonalized Hamiltonian is given by Eq.~(\ref{eq:DH}) and $D_{\AH}\left(\alpha\right)$ is computed using Eq.~(\ref{eq:DAH}). 
Denoting $x_{m}=\frac{2\pi m }{N}$ for brevity we have
\begin{IEEEeqnarray*}{rCl}
 \left[D_{\AH}\right]_{mm}&=&2\cos\left(\alpha-x_{m}\right)\\
 &=&\exp\left(i\left(\alpha-x_{m}\right)\right)+\exp\left(i\left(x_{m}-\alpha\right)\right),
\end{IEEEeqnarray*} 
and by Eq.~(\ref{eq:DH}) we find
\begin{IEEEeqnarray*}{rCl}
 N H_{mn}&=&N\sum_{l=0}^{N-1} S_{ml}\left[\DHam\right]_{ll}S^*_{ln}\\
&=&\sum_{l=0}^{N-1}\sum_{p=1}^N j_p\sum_{k=0}^p \bigl[\bigl(e^{i\left(\alpha-x_l\right)}+\>e^{-i\left(\alpha-x_l\right)}\bigr)^p\\
&&+\>\bigl(e^{i\left(-\alpha-x_l\right)}+e^{-i\left(-\alpha-x_l\right)}\bigr)^p\bigr]S_{ml} S^*_{ln}\\
&=&\sum_{l=0}^{N-1}\sum_{p=1}^Nj_p\sum_{k=0}^p\binom{p}{k}\Bigl[ e^{i\left(\alpha-x_l\right)k}e^{-i\left(\alpha-x_l\right)\left(p-k\right)}\\
&&\>+e^{-i\left(\alpha+x_l\right)k}e^{i\left(\alpha+x_l\right)\left(p-k\right)}\Bigr]e^{ix_l\left(m-n \right)}\\
&=&\sum_{l=0}^{N-1}\sum_{p=1}^Nj_p\sum_{k=0}^p{\binom{p}{k}}\,e^{ix_l\left(-2k+m-n+p\right)}\bigl[ e^{i\alpha\left(2k-p\right)}\\
&&\>+e^{-i\alpha\left(2k-p\right)}\bigr]\\
&=&\sum_{p=1}^Nj_p\!\sum_{k=0}^p{\binom{p}{k}}\,2\cos\left[\alpha\left(2k-p\right)\right]\!\sum_{l=0}^{N-1}\!
e^{ix_l\left(m-n+p-2k\right)}.
\end{IEEEeqnarray*}

Performing the $l$-sum using $\sum_{l=0}^{N-1}e^{ix_l\left(m-n\right)}=N\delta_{mn}$ we obtain Eq.~(\ref{eq:HmnCirc}):
\begin{IEEEeqnarray*}{rCl}
H_{mn}&=&\sum_{p=1}^Nj_p\sum_{k=0}^p{\binom{p}{k}}\,\delta_{2k,m-n+p}
 2\cos\left[\alpha\left(2k-p\right)\right]\\
&=&2\cos\left[\left(m-n\right)\alpha\right]\sum_{p=1}^Nj_p\sum_{k=0}^p\binom{p}{k}\,\delta_{2k,m-n+p}.
\end{IEEEeqnarray*}
Clearly, $m-n+p$ is even for terms that contribute to $H_{mn}$. 

\bibliography{paperCTQW}

\begin{thebibliography}{23}%
\makeatletter
\providecommand \@ifxundefined [1]{%
 \@ifx{#1\undefined}
}%
\providecommand \@ifnum [1]{%
 \ifnum #1\expandafter \@firstoftwo
 \else \expandafter \@secondoftwo
 \fi
}%
\providecommand \@ifx [1]{%
 \ifx #1\expandafter \@firstoftwo
 \else \expandafter \@secondoftwo
 \fi
}%
\providecommand \natexlab [1]{#1}%
\providecommand \enquote  [1]{``#1''}%
\providecommand \bibnamefont  [1]{#1}%
\providecommand \bibfnamefont [1]{#1}%
\providecommand \citenamefont [1]{#1}%
\providecommand \href@noop [0]{\@secondoftwo}%
\providecommand \href [0]{\begingroup \@sanitize@url \@href}%
\providecommand \@href[1]{\@@startlink{#1}\@@href}%
\providecommand \@@href[1]{\endgroup#1\@@endlink}%
\providecommand \@sanitize@url [0]{\catcode `\\12\catcode `\$12\catcode
  `\&12\catcode `\#12\catcode `\^12\catcode `\_12\catcode `\%12\relax}%
\providecommand \@@startlink[1]{}%
\providecommand \@@endlink[0]{}%
\providecommand \url  [0]{\begingroup\@sanitize@url \@url }%
\providecommand \@url [1]{\endgroup\@href {#1}{\urlprefix }}%
\providecommand \urlprefix  [0]{URL }%
\providecommand \Eprint [0]{\href }%
\providecommand \doibase [0]{http://dx.doi.org/}%
\providecommand \selectlanguage [0]{\@gobble}%
\providecommand \bibinfo  [0]{\@secondoftwo}%
\providecommand \bibfield  [0]{\@secondoftwo}%
\providecommand \translation [1]{[#1]}%
\providecommand \BibitemOpen [0]{}%
\providecommand \bibitemStop [0]{}%
\providecommand \bibitemNoStop [0]{.\EOS\space}%
\providecommand \EOS [0]{\spacefactor3000\relax}%
\providecommand \BibitemShut  [1]{\csname bibitem#1\endcsname}%
\let\auto@bib@innerbib\@empty
\bibitem [{\citenamefont {Shor}(1997)}]{shor94}%
  \BibitemOpen
  \bibfield  {author} {\bibinfo {author} {\bibfnamefont {P.~W.}\ \bibnamefont
  {Shor}},\ }\href {\doibase 10.1137/S0097539795293172} {\bibfield  {journal}
  {\bibinfo  {journal} {SIAM J. Comput.}\ }\textbf {\bibinfo {volume} {26}},\
  \bibinfo {pages} {1484} (\bibinfo {year} {1997})}\BibitemShut {NoStop}%
\bibitem [{\citenamefont {Grover}(1997)}]{grover97}%
  \BibitemOpen
  \bibfield  {author} {\bibinfo {author} {\bibfnamefont {L.~K.}\ \bibnamefont
  {Grover}},\ }\href {\doibase 10.1103/PhysRevLett.79.4709} {\bibfield
  {journal} {\bibinfo  {journal} {Phys. Rev. Lett.}\ }\textbf {\bibinfo
  {volume} {79}},\ \bibinfo {pages} {4709} (\bibinfo {year}
  {1997})}\BibitemShut {NoStop}%
\bibitem [{\citenamefont {Bao}\ \emph {et~al.}(2012)\citenamefont {Bao},
  \citenamefont {Xu}, \citenamefont {Li}, \citenamefont {Yuan}, \citenamefont
  {Lu},\ and\ \citenamefont {Pan}}]{Bao11122012}%
  \BibitemOpen
  \bibfield  {author} {\bibinfo {author} {\bibfnamefont {X.-H.}\ \bibnamefont
  {Bao}}, \bibinfo {author} {\bibfnamefont {X.-F.}\ \bibnamefont {Xu}},
  \bibinfo {author} {\bibfnamefont {C.-M.}\ \bibnamefont {Li}}, \bibinfo
  {author} {\bibfnamefont {Z.-S.}\ \bibnamefont {Yuan}}, \bibinfo {author}
  {\bibfnamefont {C.-Y.}\ \bibnamefont {Lu}}, \ and\ \bibinfo {author}
  {\bibfnamefont {J.-W.}\ \bibnamefont {Pan}},\ }\href {\doibase
  10.1073/pnas.1207329109} {\bibfield  {journal} {\bibinfo  {journal} {Proc.
  Natl. Acad. Sci. USA}\ }\textbf {\bibinfo {volume} {109}},\ \bibinfo {pages}
  {20347} (\bibinfo {year} {2012})}\BibitemShut {NoStop}%
\bibitem [{\citenamefont {Duan}\ \emph {et~al.}(2001)\citenamefont {Duan},
  \citenamefont {Lukin}, \citenamefont {Cirac},\ and\ \citenamefont
  {Zoller}}]{Duan2001}%
  \BibitemOpen
  \bibfield  {author} {\bibinfo {author} {\bibfnamefont {L.-M.}\ \bibnamefont
  {Duan}}, \bibinfo {author} {\bibfnamefont {M.~D.}\ \bibnamefont {Lukin}},
  \bibinfo {author} {\bibfnamefont {J.~I.}\ \bibnamefont {Cirac}}, \ and\
  \bibinfo {author} {\bibfnamefont {P.}~\bibnamefont {Zoller}},\ }\href
  {\doibase 10.1038/35106500} {\bibfield  {journal} {\bibinfo  {journal}
  {Nature}\ }\textbf {\bibinfo {volume} {414}},\ \bibinfo {pages} {413}
  (\bibinfo {year} {2001})}\BibitemShut {NoStop}%
\bibitem [{\citenamefont {Godsil}(2011)}]{Godsil2011}%
  \BibitemOpen
  \bibfield  {author} {\bibinfo {author} {\bibfnamefont {C.}~\bibnamefont
  {Godsil}},\ }\href@noop {} {\bibfield  {journal} {\bibinfo  {journal}
  {Discrete Math.}\ }\textbf {\bibinfo {volume} {312}},\ \bibinfo {pages} {129}
  (\bibinfo {year} {2011})}\BibitemShut {NoStop}%
\bibitem [{\citenamefont {Vinet}\ and\ \citenamefont
  {Zhedanov}(2012)}]{Vinet2012}%
  \BibitemOpen
  \bibfield  {author} {\bibinfo {author} {\bibfnamefont {L.}~\bibnamefont
  {Vinet}}\ and\ \bibinfo {author} {\bibfnamefont {A.}~\bibnamefont
  {Zhedanov}},\ }\href {\doibase 10.1103/PhysRevA.86.052319} {\bibfield
  {journal} {\bibinfo  {journal} {Phys. Rev. A}\ }\textbf {\bibinfo {volume}
  {86}},\ \bibinfo {pages} {052319} (\bibinfo {year} {2012})}\BibitemShut
  {NoStop}%
\bibitem [{\citenamefont {Sarovar}\ \emph {et~al.}(2010)\citenamefont
  {Sarovar}, \citenamefont {Ishizaki}, \citenamefont {Fleming},\ and\
  \citenamefont {Whaley}}]{Sarovar2010}%
  \BibitemOpen
  \bibfield  {author} {\bibinfo {author} {\bibfnamefont {M.}~\bibnamefont
  {Sarovar}}, \bibinfo {author} {\bibfnamefont {A.}~\bibnamefont {Ishizaki}},
  \bibinfo {author} {\bibfnamefont {G.~R.}\ \bibnamefont {Fleming}}, \ and\
  \bibinfo {author} {\bibfnamefont {K.~B.}\ \bibnamefont {Whaley}},\ }\href
  {\doibase 10.1038/nphys1652} {\bibfield  {journal} {\bibinfo  {journal}
  {Nature Phys.}\ }\textbf {\bibinfo {volume} {6}},\ \bibinfo {pages} {462}
  (\bibinfo {year} {2010})}\BibitemShut {NoStop}%
\bibitem [{\citenamefont {Caruso}\ \emph {et~al.}(2009)\citenamefont {Caruso},
  \citenamefont {Chin}, \citenamefont {Datta}, \citenamefont {Huelga},\ and\
  \citenamefont {Plenio}}]{caruso2009}%
  \BibitemOpen
  \bibfield  {author} {\bibinfo {author} {\bibfnamefont {F.}~\bibnamefont
  {Caruso}}, \bibinfo {author} {\bibfnamefont {A.~W.}\ \bibnamefont {Chin}},
  \bibinfo {author} {\bibfnamefont {A.}~\bibnamefont {Datta}}, \bibinfo
  {author} {\bibfnamefont {S.~F.}\ \bibnamefont {Huelga}}, \ and\ \bibinfo
  {author} {\bibfnamefont {M.~B.}\ \bibnamefont {Plenio}},\ }\href@noop {}
  {\bibfield  {journal} {\bibinfo  {journal} {J. Chem. Phys.}\ }\textbf
  {\bibinfo {volume} {131}},\ \bibinfo {eid} {105106} (\bibinfo {year}
  {2009})}\BibitemShut {NoStop}%
\bibitem [{\citenamefont {Aharonov}\ \emph {et~al.}(2001)\citenamefont
  {Aharonov}, \citenamefont {Ambainis}, \citenamefont {Kempe},\ and\
  \citenamefont {Vazirani}}]{Aharonov2001}%
  \BibitemOpen
  \bibfield  {author} {\bibinfo {author} {\bibfnamefont {D.}~\bibnamefont
  {Aharonov}}, \bibinfo {author} {\bibfnamefont {A.}~\bibnamefont {Ambainis}},
  \bibinfo {author} {\bibfnamefont {J.}~\bibnamefont {Kempe}}, \ and\ \bibinfo
  {author} {\bibfnamefont {U.}~\bibnamefont {Vazirani}},\ }in\ \href {\doibase
  10.1145/380752.380758} {\emph {\bibinfo {booktitle} {Proc. of the 33rd Annual
  ACM Symposium on Theory of Computing}}},\ \bibinfo {series and number} {STOC
  '01}\ (\bibinfo  {publisher} {ACM},\ \bibinfo {address} {New York, NY, USA},\
  \bibinfo {year} {2001})\ pp.\ \bibinfo {pages} {50--59}\BibitemShut {NoStop}%
\bibitem [{\citenamefont {Farhi}\ and\ \citenamefont
  {Gutmann}(1998)}]{PhysRevA.58.915}%
  \BibitemOpen
  \bibfield  {author} {\bibinfo {author} {\bibfnamefont {E.}~\bibnamefont
  {Farhi}}\ and\ \bibinfo {author} {\bibfnamefont {S.}~\bibnamefont
  {Gutmann}},\ }\href {\doibase 10.1103/PhysRevA.58.915} {\bibfield  {journal}
  {\bibinfo  {journal} {Phys. Rev. A}\ }\textbf {\bibinfo {volume} {58}},\
  \bibinfo {pages} {915} (\bibinfo {year} {1998})}\BibitemShut {NoStop}%
\bibitem [{\citenamefont {M{\"u}lken}\ and\ \citenamefont
  {Blumen}(2011)}]{Mulken2011}%
  \BibitemOpen
  \bibfield  {author} {\bibinfo {author} {\bibfnamefont {O.}~\bibnamefont
  {M{\"u}lken}}\ and\ \bibinfo {author} {\bibfnamefont {A.}~\bibnamefont
  {Blumen}},\ }\href {\doibase http://dx.doi.org/10.1016/j.physrep.2011.01.002}
  {\bibfield  {journal} {\bibinfo  {journal} {Phys. Rep.}\ }\textbf {\bibinfo
  {volume} {502}},\ \bibinfo {pages} {37 } (\bibinfo {year}
  {2011})}\BibitemShut {NoStop}%
\bibitem [{\citenamefont {Venegas-Andraca}(2012)}]{VenegasAndraca:2012fh}%
  \BibitemOpen
  \bibfield  {author} {\bibinfo {author} {\bibfnamefont {S.~E.}\ \bibnamefont
  {Venegas-Andraca}},\ }\href {\doibase 10.1007/s11128-012-0432-5} {\bibfield
  {journal} {\bibinfo  {journal} {Quant. Inf. Proc.}\ }\textbf {\bibinfo
  {volume} {11}},\ \bibinfo {pages} {1015} (\bibinfo {year}
  {2012})}\BibitemShut {NoStop}%
\bibitem [{\citenamefont {{Childs}}(2010)}]{Childs2010discCont}%
  \BibitemOpen
  \bibfield  {author} {\bibinfo {author} {\bibfnamefont {A.~M.}\ \bibnamefont
  {{Childs}}},\ }\href {\doibase 10.1007/s00220-009-0930-1} {\bibfield
  {journal} {\bibinfo  {journal} {Commun. Math. Phys.}\ }\textbf {\bibinfo
  {volume} {294}},\ \bibinfo {pages} {581} (\bibinfo {year}
  {2010})}\BibitemShut {NoStop}%
\bibitem [{\citenamefont {{Childs}}(2009)}]{Childs2009quantalgo}%
  \BibitemOpen
  \bibfield  {author} {\bibinfo {author} {\bibfnamefont {A.~M.}\ \bibnamefont
  {{Childs}}},\ }\href {\doibase 10.1103/PhysRevLett.102.180501} {\bibfield
  {journal} {\bibinfo  {journal} {Phys. Rev. Lett.}\ }\textbf {\bibinfo
  {volume} {102}},\ \bibinfo {eid} {180501} (\bibinfo {year}
  {2009})}\BibitemShut {NoStop}%
\bibitem [{\citenamefont {Lovett}\ \emph {et~al.}(2010)\citenamefont {Lovett},
  \citenamefont {Cooper}, \citenamefont {Everitt}, \citenamefont {Trevers},\
  and\ \citenamefont {Kendon}}]{PhysRevA.81.042330}%
  \BibitemOpen
  \bibfield  {author} {\bibinfo {author} {\bibfnamefont {N.~B.}\ \bibnamefont
  {Lovett}}, \bibinfo {author} {\bibfnamefont {S.}~\bibnamefont {Cooper}},
  \bibinfo {author} {\bibfnamefont {M.}~\bibnamefont {Everitt}}, \bibinfo
  {author} {\bibfnamefont {M.}~\bibnamefont {Trevers}}, \ and\ \bibinfo
  {author} {\bibfnamefont {V.}~\bibnamefont {Kendon}},\ }\href {\doibase
  10.1103/PhysRevA.81.042330} {\bibfield  {journal} {\bibinfo  {journal} {Phys.
  Rev. A}\ }\textbf {\bibinfo {volume} {81}},\ \bibinfo {pages} {042330}
  (\bibinfo {year} {2010})}\BibitemShut {NoStop}%
\bibitem [{\citenamefont {Zimbor\'{a}s}\ \emph {et~al.}(2013)\citenamefont
  {Zimbor\'{a}s}, \citenamefont {Faccin}, \citenamefont {K\'{a}d\'{a}r},
  \citenamefont {Whitfield}, \citenamefont {Lanyon},\ and\ \citenamefont
  {Biamonte}}]{zimboras13}%
  \BibitemOpen
  \bibfield  {author} {\bibinfo {author} {\bibfnamefont {Z.}~\bibnamefont
  {Zimbor\'{a}s}}, \bibinfo {author} {\bibfnamefont {M.}~\bibnamefont
  {Faccin}}, \bibinfo {author} {\bibfnamefont {Z.}~\bibnamefont
  {K\'{a}d\'{a}r}}, \bibinfo {author} {\bibfnamefont {J.}~\bibnamefont
  {Whitfield}}, \bibinfo {author} {\bibfnamefont {B.}~\bibnamefont {Lanyon}}, \
  and\ \bibinfo {author} {\bibfnamefont {J.}~\bibnamefont {Biamonte}},\
  }\href@noop {} {\bibfield  {journal} {\bibinfo  {journal} {Sci. Rep.}\
  }\textbf {\bibinfo {volume} {3}},\ \bibinfo {pages} {2361} (\bibinfo {year}
  {2013})}\BibitemShut {NoStop}%
\bibitem [{\citenamefont {Lloyd}(1996)}]{10.2307/2899535}%
  \BibitemOpen
  \bibfield  {author} {\bibinfo {author} {\bibfnamefont {S.}~\bibnamefont
  {Lloyd}},\ }\href {http://www.jstor.org/stable/2899535} {\bibfield  {journal}
  {\bibinfo  {journal} {Science}\ }\textbf {\bibinfo {volume} {273}},\ \bibinfo
  {pages} {1073} (\bibinfo {year} {1996})}\BibitemShut {NoStop}%
\bibitem [{Note1()}]{Note1}%
  \BibitemOpen
  \bibinfo {note} {As a consequence of the Cayley-Hamilton theorem, power
  series of matrices (or vector space endomorphisms) form an $N$-dimensional
  vector space and can thus be expressed as a polynomial in $A_{\protect
  \mathrm {H}}$ of degree at most $N-1$.}\BibitemShut {Stop}%
\bibitem [{\citenamefont {Gray}(2006)}]{Gray2006}%
  \BibitemOpen
  \bibfield  {author} {\bibinfo {author} {\bibfnamefont {R.~M.}\ \bibnamefont
  {Gray}},\ }\href@noop {} {\bibfield  {journal} {\bibinfo  {journal} {Found.
  Trends Commun. Inf. Theory}\ }\textbf {\bibinfo {volume} {2}},\ \bibinfo
  {pages} {155} (\bibinfo {year} {2006})}\BibitemShut {NoStop}%
\bibitem [{\citenamefont {Tsomokos}\ \emph {et~al.}(2008)\citenamefont
  {Tsomokos}, \citenamefont {Plenio}, \citenamefont {de~Vega},\ and\
  \citenamefont {Huelga}}]{tsomokos2008state}%
  \BibitemOpen
  \bibfield  {author} {\bibinfo {author} {\bibfnamefont {D.~I.}\ \bibnamefont
  {Tsomokos}}, \bibinfo {author} {\bibfnamefont {M.~B.}\ \bibnamefont
  {Plenio}}, \bibinfo {author} {\bibfnamefont {I.}~\bibnamefont {de~Vega}}, \
  and\ \bibinfo {author} {\bibfnamefont {S.~F.}\ \bibnamefont {Huelga}},\
  }\href {\doibase 10.1103/PhysRevA.78.062310} {\bibfield  {journal} {\bibinfo
  {journal} {Phys. Rev. A}\ }\textbf {\bibinfo {volume} {78}},\ \bibinfo
  {pages} {062310} (\bibinfo {year} {2008})}\BibitemShut {NoStop}%
\bibitem [{\citenamefont {Solenov}\ and\ \citenamefont
  {Fedichkin}(2006)}]{PhysRevA.73.012313}%
  \BibitemOpen
  \bibfield  {author} {\bibinfo {author} {\bibfnamefont {D.}~\bibnamefont
  {Solenov}}\ and\ \bibinfo {author} {\bibfnamefont {L.}~\bibnamefont
  {Fedichkin}},\ }\href {\doibase 10.1103/PhysRevA.73.012313} {\bibfield
  {journal} {\bibinfo  {journal} {Phys. Rev. A}\ }\textbf {\bibinfo {volume}
  {73}},\ \bibinfo {pages} {012313} (\bibinfo {year} {2006})}\BibitemShut
  {NoStop}%
\bibitem [{\citenamefont {Guy}\ and\ \citenamefont
  {Harary}(1967)}]{moebius-ladder}%
  \BibitemOpen
  \bibfield  {author} {\bibinfo {author} {\bibfnamefont {R.~K.}\ \bibnamefont
  {Guy}}\ and\ \bibinfo {author} {\bibfnamefont {F.}~\bibnamefont {Harary}},\
  }\href@noop {} {\bibfield  {journal} {\bibinfo  {journal} {Canad. Math.
  Bull.}\ }\textbf {\bibinfo {volume} {10}},\ \bibinfo {pages} {493} (\bibinfo
  {year} {1967})}\BibitemShut {NoStop}%
\bibitem [{\citenamefont {Newman}(2010)}]{Newman_2010}%
  \BibitemOpen
  \bibfield  {author} {\bibinfo {author} {\bibfnamefont {M.~E.~J.}\
  \bibnamefont {Newman}},\ }\href@noop {} {\emph {\bibinfo {title} {Networks:
  An Introduction}}}\ (\bibinfo  {publisher} {Oxford University Press},\
  \bibinfo {year} {2010})\BibitemShut {NoStop}%
\end{thebibliography}%
\end{document}